\documentclass{article}
\usepackage[T1]{fontenc} 
\usepackage[utf8]{inputenc} 
\usepackage{ismir,amsmath,cite,url}
\usepackage{graphicx}
\usepackage{color}

\usepackage{lineno}

\title{EditGen: Harnessing Cross Attention Control for Instruction-Based auto-regressive Audio Editing}

\author{
Vassilis Sioros\\
  \small{Department of Informatics and Telecommunications, UoA} \\
  \small{NCSR "Demokritos"} \\
  \small{\tt {vsioros@di.uoa.gr}}
  \and
  Alexandros Potamianos\\
  \small{School of Electrical and Computer Engineering, NTUA} \\
  \small{\tt {potam@central.ntua.gr}}
  \and
  Giorgos Paraskevopoulos\\
  \small{School of Electrical and Computer Engineering, NTUA} \\
  \small{\tt {geopar@central.ntua.gr}}
}

\def\authorname{F. Author, S. Author, and T. Author}

\usepackage[bookmarks=false,pdfauthor={\authorname},pdfsubject={\papersubject},hidelinks]{hyperref}

\begin{document}

\maketitle
\begin{abstract}
In this study, we investigate leveraging cross-attention control for efficient audio editing within auto-regressive models. Inspired by image editing methodologies, we develop a Prompt-to-Prompt-like approach that guides edits through cross and self-attention mechanisms. Integrating a diffusion-based strategy, influenced by Auffusion, we extend the model's functionality to support refinement edits, establishing a baseline for prompt-guided audio editing. Additionally, we introduce an alternative approach by incorporating MUSICGEN, a pre-trained frozen auto-regressive model, and propose three editing mechanisms, based on Replacement, Reweighting, and Refinement of the attention scores. We employ commonly-used music-specific evaluation metrics and a human study, to gauge time-varying controllability, adherence to global text cues, and overall audio realism. The automatic and human evaluations indicate that the proposed combination of prompt-to-prompt guidance with autoregressive generation models significantly outperforms the diffusion-based baseline in terms of melody, dynamics, and tempo of the generated audio.
\end{abstract}
\section{Introduction}\label{sec:introduction}

\begin{figure}
 \centerline{
 \includegraphics[width=0.9\columnwidth]{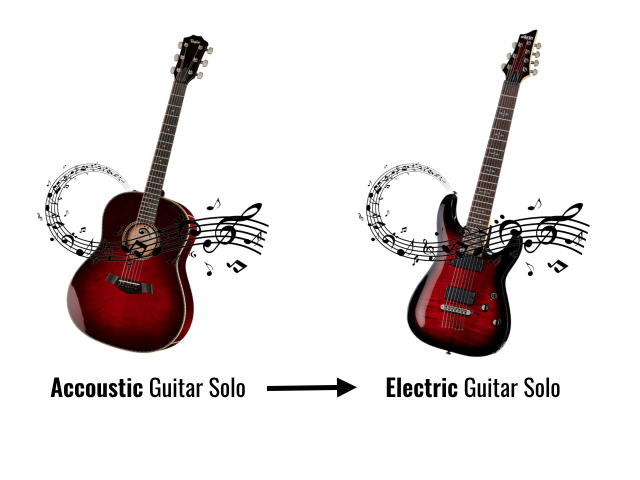}}
 \caption{Task Overview: By leveraging cross-attention layers, we perform audio tasks such as token value adjustments, global audio editing, and amplification/attenuation of semantic effects without the need for model retraining.}
 \label{fig:introduction}
\end{figure}

Obtaining satisfactory outcomes in audio manipulation tasks has typically required large datasets with intricate annotations, a process that is often labor-intensive and costly. Moreover, crafting effective model architectures tailored to the nuances of audio processing requires substantial expertise and experimentation. Fine-tuning existing model architectures, while a possibility, remains a costly endeavor. 

In this work, we employ prompt-guided models in the audio domain, and, specifically, we adapt the Prompt-to-Prompt \cite{hertz_prompt--prompt_2022} technique, previously successful in image manipulation, to audio editing. This approach allows for fine-grained, semantically meaningful audio editing without the need for model retraining or additional data. Prompt-to-Prompt leverages cross-attention maps to control the generation process, enabling users to influence how generated data units interact with tokens. This versatility enables tasks such as token value changes, global audio editing, and the manipulation of semantic effects without modifying the underlying model.

We apply Prompt-to-Prompt for the task of music editing and adapt it for an auto-regressive prompt-based model. This marks the first successful use of prompt-to-prompt in the auto-regressive model audio editing context. Our extensive evaluation demonstrates these techniques' potential for intuitive text-based audio editing. Our key contributions are:

\begin{enumerate}
    \item We explore the utilization of a pre-trained frozen auto-regressive transformer model, initially designed for generating high-quality music samples from a given text prompt, for audio editing.
    \item We design and implement three distinct audio editing mechanisms, inspired by those presented in \cite{hertz_prompt--prompt_2022}.
    \item We evaluate our auto-regressive transformer model-based approach against existing diffusion-based methods using automatic music-related metrics and feedback from users.
\end{enumerate}


\section{Related Work}\label{sec:related_work}


\textbf{Diffusion-based audio generation:} Diffusion-based models have been widely explored for generation tasks in the audio and music domains. Yang et al. \cite{yang_diffsound_2023} employ a VQ-VAE model trained on mel-spectrograms to convert them into discrete codes. These codes are then fed into a diffusion model to generate audio signals. Make-An-Audio \cite{huang_make--audio_2023} employs a spectrogram autoencoder, integrates CLAP \cite{elizalde_clap_2022} and introduces a pseudo prompt enhancement approach, aligning natural languages with audio data, enabling utilization of vast unsupervised language-free data. AudioLDM \cite{liu_audioldm_2023-3} employs a latent diffusion model (LDM) and addresses the limitations of paired data methods by training generative models exclusively with audio data. AudioLDM 2 \cite{liu_audioldm_2023-1} introduces the Language of Audio (LOA), employing AudioMAE \cite{huang_masked_2023} pre-trained on diverse audio content. AUDIT \cite{wang_audit_2023} combines latent diffusion models with human instructions to generate edited audio segments using both audio and text cues. With regards to music, TANGO \cite{ghosal_text--audio_2023}, inspired by latent diffusion models (LDM) and AudioLDM, utilizes an LLM instead of CLAP-based embeddings. Auffusion\cite{xue_auffusion_2024} employs a pretrained Latent Diffusion Model (LDM) and HiFi-GAN \cite{kong_hifi-gan_2020} vocoder. Additionally, the authors introduce a cross-attention mechanism that enhances alignment and flexibility. MusicLDM \cite{chen_musicldm_2023}, adapts Stable Diffusion and AudioLDM architectures to the music domain. To address the challenge of limited training data and promote innovation, novel mixup strategies are proposed: beat-synchronous audio mixup (BAM) and beat-synchronous latent mixup (BLM). InstructME \cite{han_instructme_2023} leverages a tailored latent diffusion model, facilitating tasks such as adding, removing, and remixing musical elements while preserving harmonic integrity via chord progression matrices.

\vspace*{1mm}
\noindent
\textbf{Auto-regressive audio generation:} As an alternative to diffusion-based models for audio and music generation, autoregressive models have shown promise in recent years. WaveNet \cite{oord_wavenet_2016} introduced an auto-regressive classification method for speech synthesis, surpassing traditional concatenative and parametric approaches, albeit with slower inference. AudioGen \cite{kreuk_audiogen_2023}, has outperformed Diffsound by employing auto-regressive modeling in discrete waveform spaces. Jukebox \cite{dhariwal_jukebox_2020} uses a multi-scale VQ-VAE \cite{oord_neural_2018} to compress raw audio into discrete codes, which are then modeled using auto-regressive transformers. AudioLM \cite{borsos_audiolm_2023} utilizes tokens generated by a SoundStream \cite{zeghidour_soundstream_2021} neural codec \cite{kankanahalli_end--end_2021,petermann_harp-net_2021} as targets for a sequence modeling task. In \cite{agostinelli_musiclm_2023}, Agostinelli et al. introduce MusicLM, following a similar approach to AudioLM but tailored for music editing tasks. MUSICGEN, a transformer-based decoder introduced in \cite{copet_simple_2023}, employs EnCodec \cite{defossez_high_2022} for audio tokenization. Text conditioning integrates techniques like T5 encoder, FLAN-T5, and CLAP, while melody conditioning utilizes dominant time-frequency bins to control melodic structure. The model also introduces a framework for codebook interleaving patterns, improving efficiency and flexibility.

\vspace*{1mm}
\noindent
\textbf{Editing techniques:} Prompt-to-Prompt \cite{hertz_prompt--prompt_2022} leverages internal cross-attention layers to control which pixels attend to which tokens, allowing tasks like token value changes, global image editing, and semantic effects amplification/attenuation without model retraining. Textual Inversion \cite{mokady_null-text_2022} utilizes a sequence of noised latent codes obtained from initial DDIM inversion as a pivot and optimizes the null-text embedding. By fine-tuning a pre-trained text-to-image model with a few images of a subject, Dreambooth \cite{ruiz_dreambooth_2023} associates a unique identifier with the object. Leveraging the semantic prior embedded in the model along with a new autogenous class-specific prior preservation loss, Dreambooth enables the synthesis of photorealistic images of the subject contextualized in various scenes. In \cite{kumari_multi-concept_2023} the authors proposed Custom Diffusion, where they optimize only a few parameters in the text-to-image conditioning mechanism to represent new concepts. This approach performs on par with or better than existing methods while maintaining computational efficiency. SVDiff \cite{han_svdiff_2023}, fine-tunes the singular values of the weight matrices resulting in a compact parameter space, reducing the risk of overfitting and language-drifting \cite{lee_countering_2019}. \cite{plitsis_investigating_2023} employs Textual Inversion \cite{mokady_null-text_2022} and Dreambooth \cite{ruiz_dreambooth_2023} to personalize the outputs of AudioLDM for newly learned musical concepts in a few-shot manner. In \cite{manor_zero-shot_2024}, the authors investigate two zero-shot audio editing techniques utilizing DDPM inversion on pre-trained diffusion networks. Their approach, based on \cite{huberman-spiegelglas_edit_2024}, involves extracting latent noise vectors corresponding to the source signal and using these vectors in a DDPM sampling process to guide the diffusion toward the desired edit. For text-based editing, they adjust the text prompt given to the denoiser model. In the unsupervised scenario, they perturb the denoiser output along the directions of the top principal components (PCs) of the posterior.

\vspace*{1mm}
Building upon the achievements of Prompt-to-Prompt and the superior audio quality of MUSICGEN, we aim to integrate them into an auto-regressive framework. This will blend Prompt-to-Prompt's editability with the high-quality sound of modern auto-regressive models like MUSICGEN.

\section{Methodology}\label{sec:methodology}

\subsection{MUSICGEN}

MUSICGEN \cite{copet_simple_2023} employs EnCodec \cite{defossez_high_2022}, a convolutional auto-encoder utilizing Residual Vector Quantization (RVQ) for latent space quantization. The input, a reference audio random variable $X$, is encoded into a continuous tensor with a lower frame rate ($f_r$) compared to the sample rate ($f_s$). The continuous representation is then quantized into discrete tokens ($Q$) using RVQ, resulting in $K$ parallel sequences (for each time step), each with $T$ tokens, where $K$ is the number of codebooks, and $M$ is the codebook size. The authors employ an autoregressive decomposition approach that predicts multiple codebooks simultaneously, and thusly greatly accelerates both training and inference. More specifically, MusicGen employs a token interleaving pattern to generate all codebooks in a single decoder pass, eliminating the need for cascading multiple models and making inference much faster.

\subsection{Prompt-to-Prompt}

Consider an audio sample \( A \) generated using a text prompt \( P \). By injecting the attention maps obtained during \( A \)'s generation into a new generation with a modified prompt \( P^* \), we can perform a meaningful edit resulting in a new audio sample \( A^* \) that preserves the original's structure. To address specific editing operations, we employ three editing mechanisms akin to the ones introduced in \cite{hertz_prompt--prompt_2022}:

\vspace*{2mm}
\noindent
\textbf{Replace:} The user swaps tokens of the original prompt with others. For example replacing an acoustic with an electric guitar (\figref{fig:introduction}). We inject the attention maps of the source sample into the generation process with the modified prompt:

\begin{equation}
    Edit(M_t, M^*_t, t) = \left\{
\begin{array}{ll}
      M^*_t, & \text{if } t < \tau \\
      M_t, & otherwise \\
\end{array} 
\right. 
\end{equation}
where $\tau$ is a timestamp parameter that determines until which step the injection is applied.

\vspace*{2mm}
\noindent
\textbf{Refine:} The user adds new tokens to the prompt. In this case, attention injection is applied only to the common tokens shared by both prompts. Formally, we utilize an alignment function $A$ that takes a token index from the target prompt $P^*$ and outputs the corresponding token index in the original prompt $P$ or $\emptyset$ if there isn't a match:

\begin{equation}
    (Edit(M_t, M^*_t, t))_{i, j} = \left\{
\begin{array}{ll}
      (M^*_t)_{i,j}, & \text{if } A(j) = \emptyset \\
      (M_t)_{i, A(j)}, & otherwise \\
\end{array} 
\right. 
\end{equation}
It's worth recalling that index $i$ corresponds to a value, while index $j$ corresponds to a text token. Once again, we may set a timestamp $\tau$ to control the number of steps in which the injection is applied.

\vspace*{2mm}
\noindent
\textbf{Reweight:} Finally, the user may wish to strengthen or weaken the extent to which each token affects the result. To achieve this manipulation, we scale the attention map of the assigned token $j^*$ with a parameter $c$ ranging from -2 to 2, resulting in a stronger or weaker effect. The attention maps for the other tokens remain unchanged:

\begin{equation}
    (Edit(M_t, M^*_t, t))_{i, j} = \left\{
\begin{array}{ll}
      c \cdot (M_t)_{i, j}, & \text{if } j = j* \\
      (M_t)_{i, j}, & otherwise \\
\end{array} 
\right. 
\end{equation}

\vspace*{2mm}
In the original implementation of Prompt-to-Prompt, utilizing text-guided diffusion models \cite{saharia_photorealistic_2022}, the output was decided early in the diffusion process. By restricting the number of injection steps $\tau$, the authors managed to steer the generation process while maintaining flexibility in adapting the geometry to the new prompt. To align Prompt-to-Prompt with MUSICGEN's auto-regressive features, we apply the attention injection procedure at all timesteps. Since MUSICGEN treats audio generation as a sequence-to-sequence task, the notion of time doesn't correspond to the application of an iterative method like in the case of diffusion models, but rather to sampling new audio tokens. Thus, to guarantee that edits affect the entirety of the generated audio, this adjustment was deemed essential. As mentioned earlier, edits in the context of the original Prompt-to-Prompt paper, tailored for diffusion models, were applied for a set number of iterations of the diffusion process. This strategy aimed to strike a balance between generating novel samples and preserving the essential characteristics of the original. Acknowledging the auto-regressive nature of MUSICGEN, we investigate an alternative method: soft-blending. This technique merges the feature maps generated during the forward process with the injected ones, employing a weighted average for the output. Let \( X_{t} \) denote the feature map generated at time step \( t \) during the forward process, and let \( Y_{t} \) represent the injected feature map. The soft-blending technique combines these feature maps using a weighted average to produce the output feature map \( Z_{t} \):

\begin{equation}
   Z_{t} = \alpha X_{t} + (1 - \alpha) Y_{t} 
\end{equation}
where \( \alpha \) is the blending parameter, with values between 0 and 1, and is expressed as:

\[ \alpha = \frac{i}{N} \]

\noindent
where \( i \) represents the index of the current attention layer being considered and \( N \) represents the total number of attention layers in the decoder stack. This formulation ensures that the blending factor dynamically adjusts based on the position within the decoder stack and replicates the original diffusion-based approach of prompt-to-prompt, where edits are applied for a set number of diffusion iterations.

\section{Experimental setup}

\subsection{Dataset construction}

To evaluate our method, we initially create a dataset containing prompt pairs for each editing mechanism: Replace, Refine, and Reweight. Each pair consists of original and edited text prompts. Creating our dataset consists of two steps: (1) handpicking a small group of prompt pairs, followed by (2) leveraging ChatGPT 3.5 to generate additional pairs. This hybrid approach ensures a blend of manually crafted prompts and dynamically generated ones, providing a diverse range of inputs for evaluation. We consider different audio editing axes to organize our dataset effectively. Each axis represents a distinct aspect of audio content, offering unique opportunities for creative expression and artistic exploration:

    \noindent \textbf{Instrument Change}: Substituting one instrument or sound source with another, enhancing the audio by adding nuanced details, or recalibrating the emphasis on various sonic elements. This axis enables exploring diverse timbres, textures, and sonic characteristics within the audio composition.
    
    \noindent \textbf{Mood/Tonal Change}: Mood/tonal change involves altering, changing, or enhancing the emotional resonance and tonal nuances of the music. This axis encompasses modifications that evoke different emotional responses or shift the overall tonal coloration of the audio material.
    
    \noindent \textbf{Genre Shift}: Genre shift entails transitioning between different musical styles or genres. This axis facilitates the exploration of diverse stylistic conventions, rhythmic patterns, and instrumental arrangements across various musical genres. Genre shifts offer opportunities for creative experimentation and genre-blending.

    \vspace*{1mm}
    \noindent \textbf{Melodic Transformation}: Melodic transformation involves altering the melodic content of the music. This axis encompasses modifications to melodic contours, intervals, motifs, and themes, allowing for creative reinterpretations and variations of melodic material.
    
    \noindent \textbf{Harmonic Modification}: This axis encompasses changes in chord progressions, harmonic rhythm, harmonic density, and harmonic tension, allowing for harmonic enrichment and exploration of tonal relationships. By exploring harmonic modifications, we can investigate how changes in harmonic progressions, chord voicings, and harmonic textures influence the harmonic character and emotional resonance of the music.
    
    \noindent \textbf{Form/Structure Variation}: Form/structure variation involves variations in the overall form or structure of the music. This axis encompasses changes in sectional arrangement, repetitions, transitions, and developmental processes, allowing for structural experimentation and narrative exploration.

Using MUSICGEN and Prompt-to-Prompt (as defined for autoregressive models), we generated 22 samples per edit category (Replace, Refine, and Reweight) with 5 random seeds per prompt pair. This process was repeated for Auffusion resulting in a total of 660 generated samples across both models.

\subsection{Automated music coherence evaluation metrics}


We employ multiple common evaluation metrics to assess the musical characteristics of the generated samples, including time-varying controllability, adherence to global text control, and overall audio realism:

\vspace*{1mm}
  \noindent \textbf{Melody Accuracy}: Assesses the alignment of pitch classes (C, C\#, ..., B; totaling 12) on a frame-by-frame basis between the source audio and the one derived from the application of Prompt-to-Prompt \cite{wu2023music}.
  
\vspace*{1mm}
  \noindent \textbf{Dynamics Correlation}: Refers to Pearson’s correlation between the source dynamics values on a frame-by-frame basis and the values derived from the application of Prompt-to-Prompt \cite{wu2023music}.

  
  \vspace*{1mm}
\noindent \textbf{Rhythm F1 Score}: Adheres to the conventional approach to detecting beats and downbeats \cite{mckinney_evaluation_2007,raffel_mir_eval_2014}, measuring the synchronization between the estimated timestamps of beats/downbeats derived from the source rhythm control and those generated from the application of Prompt-to-Prompt. Timestamps are obtained by applying an HMM postfilter \cite{krebs_efficient_2015} to the frame-wise probabilities of beats/downbeats (i.e., the rhythm control signal). Following \cite{raffel_mir_eval_2014}, alignment between input and generated timestamps is considered if their difference is less than 70 milliseconds.
  
  \vspace*{1mm}
\noindent \textbf{CLAP Score}: \cite{wu_large-scale_2024,chen_hts-at_2022} assesses text control adherence by calculating the cosine similarity between text and audio embeddings extracted from the CLAP model. CLAP is a dual-encoder foundation model with separate encoders for text and audio inputs. These encoders learn embedding spaces through a contrastive objective \cite{oord_representation_2019}.
  

\section{Results}\label{sec:results}

\begin{figure}
 \centerline{
 \includegraphics[width=0.9\columnwidth]{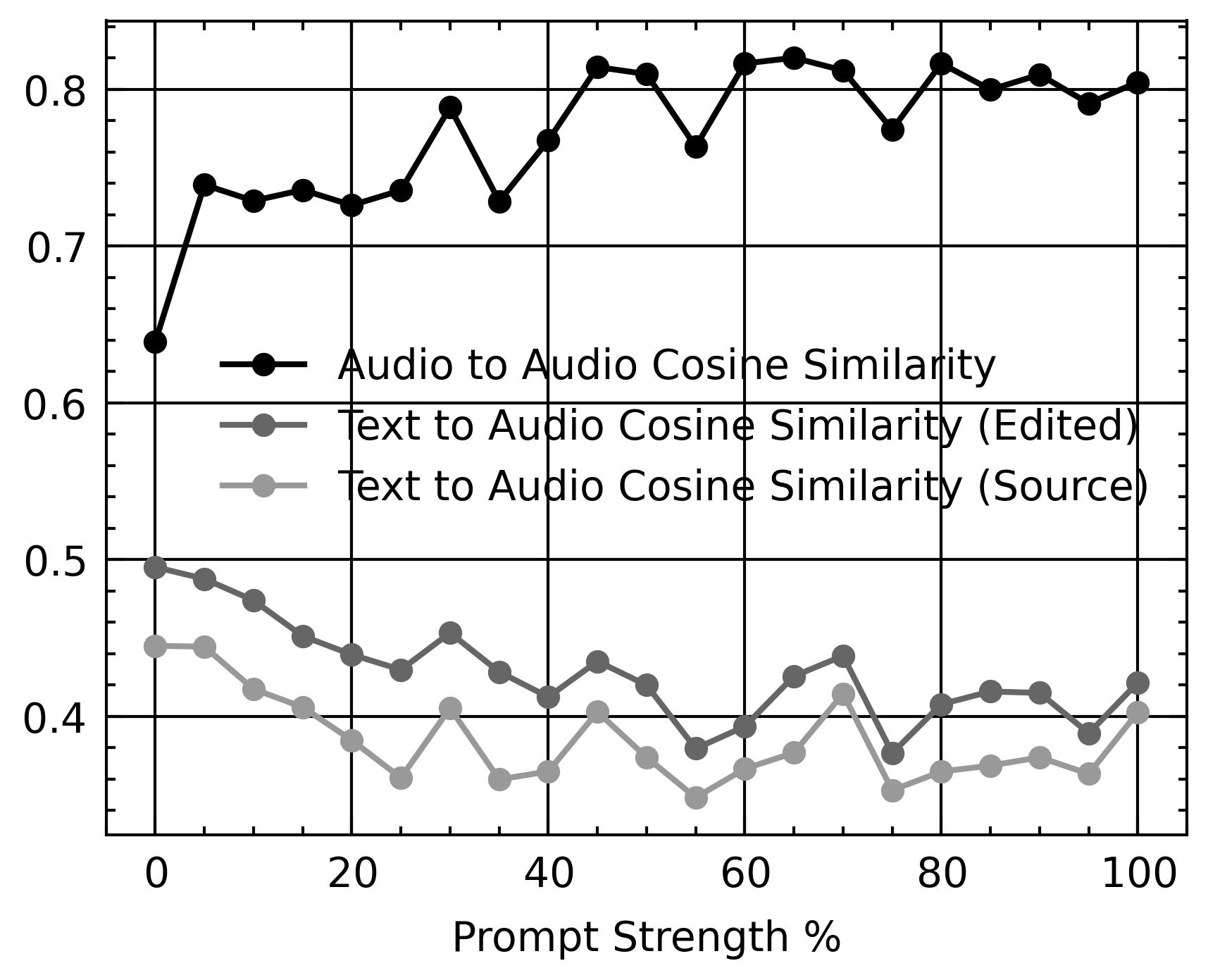}}
 \caption{Evaluation of audio and textual alignment with regards to \textit{"prompt strength"}.}
 \label{fig:results:preliminary:prompt_strength}
\end{figure}

We perform an initial experiment where we systematically vary the impact -\textit{"prompt strength"}- of injected attention maps in audio generation, gradually increasing the influence of textual cues on editing. We calculate the average cosine similarity between original and edited prompts in both Audio-to-Audio and Text-to-Audio contexts. \figref{fig:results:preliminary:prompt_strength} shows that the edited audio maintains the qualities of the original, as indicated by the high mean Audio-to-Audio cosine similarity and the mean Text-to-Audio cosine similarity with the source prompt, which remains consistent regardless of the \textit{"prompt strength"}. Additionally, the mean Text-to-Audio cosine similarity with the edited prompt, which remains stable, highlights that the edited audio remains in alignment with the edited prompt regardless of \textit{"prompt strength"}.

\subsection{The effect of soft-blending}

Additionally, we evaluate the effectiveness of soft blending cross-attention features quantitatively by computing Audio-to-Audio and Text-to-Audio cosine similarity metrics between the generated audio samples and the edited prompt. Audio-to-Audio and Text-to-Audio cosine similarity scores are averaged across the dataset, ensuring a comprehensive and objective evaluation. As indicated by \tabref{table:results:self-attn}, using soft-blending leads to higher mean Audio-to-Audio and Text-to-Audio Cosine Similarity scores. Notably, the standard deviation remains relatively the same, highlighting the reliability and consistency of the soft-blending method.

\begin{table}[t]
 \begin{center}
 \begin{tabular}{|c|c|c|}
  \hline
  \textbf{Configuration} & \textbf{T2A Similarity} & \textbf{A2A Similarity} \\
  \hline
  Hard-blending & $0.836 \mp 0.087$ & $0.400 \mp 0.152$ \\
  \hline
  Soft-blending & $0.849 \mp 0.094$ & $0.414 \mp 0.157$ \\
  \hline
 \end{tabular}
\end{center}
 \caption{Text-to-Audio \& Audio-to-Audio Cosine Similarity with regards to blending strategy.}
 \label{table:results:self-attn}
\end{table}

\subsection{Comparison of automated music metrics}

Finally, we seek to investigate the effectiveness of Prompt-to-Prompt techniques in audio editing by examining both diffusion-based and auto-regressive models. This exploration aims to offer insights into the strengths and limitations of these models for creative audio manipulation. Our baseline strategy employs Auffusion \cite{xue_auffusion_2024}, a diffusion-based approach. This method seamlessly integrates with the Prompt-to-Prompt methodology, providing a solid foundation for exploring prompt-guided audio editing. Additionally, we introduce an alternative avenue by incorporating a pre-trained frozen auto-regressive model, capitalizing on the advanced capabilities offered by the state-of-the-art MUSICGEN \cite{copet_simple_2023} model.

\begin{figure}
 \centerline{
 \includegraphics[width=0.9\columnwidth]{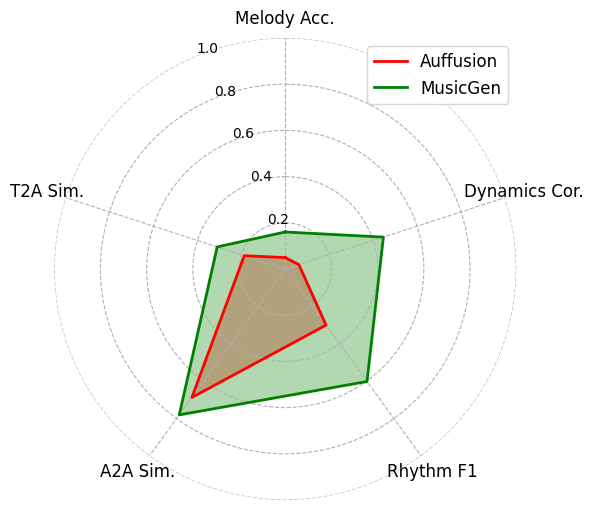}}
 \caption{Average evaluation metrics across editing mechanisms for both Auffusion and MUSICGEN.}
 \label{fig:results:comparison:spider}
\end{figure}

The metrics are averaged across the entire dataset and presented in \figref{fig:results:comparison:spider} while detailed results are showcased on \tabref{table:results:human:eval:faithfulness}. From Melody Accuracy to Dynamics Correlation, Rhythm F1 score, Audio-to-Audio cosine similarity, and Text-to-Audio cosine similarity, each metric provides a unique perspective on the capabilities of MUSICGEN and Auffusion in handling different audio editing tasks. Based on the provided data, MUSICGEN outperforms Auffusion across all evaluation metrics. It excels in capturing melody accuracy, exhibits superior similarity to both the original audio and target text prompt, and significantly outperforms Auffusion in dynamics correlation and rhythm F1 score. Our methodology, utilizing prompt-to-prompt in the auto-regressive model context for audio editing tasks, outperforms Auffusion with the MUSICGEN model. Notably, this marks the first successful use of prompt-to-prompt in the auto-regressive model audio editing context, highlighting its significance in achieving superior results.

\begin{table}[t]
\resizebox{\columnwidth}{!}{
 \begin{centering}

 \begin{tabular}{|c|c|c|c|c|c|}
  \hline
  \textbf{Edit} & \textbf{Model} & \textbf{Alignment} & \textbf{Dynamics} & \textbf{Melody} & \textbf{Tempo} \\
  \hline
  Refine & Auffusion & $2.76$ & $2.57$ & $2.41$ & $2.81$ \\
  \hline
  Refine & MUSICGEN & $\mathbf{3.43}$ & $\mathbf{3.08}$ & $\mathbf{3.22}$ & $\mathbf{3.62}$ \\
  \hline
  Replace & Auffusion & $2.91$ & $2.80$ & $2.44$ & $2.74$ \\
  \hline
  Replace & MUSICGEN & $\mathbf{3.05}$ & $\mathbf{3.07}$ & $\mathbf{3.16}$ & $\mathbf{3.56}$ \\
  \hline
  Reweight & Auffusion & $2.78$ & $2.88$ & $2.58$ & $2.78$ \\
  \hline
  Reweight & MUSICGEN & $\mathbf{3.54}$ & $\mathbf{3.24}$ & $\mathbf{3.38}$ & $\mathbf{3.86}$ \\
  \hline
 \end{tabular}
 \end{centering}
}
 \caption{Comparison of audio editing capabilities of MUSICGEN and Auffusion based on MOS (Mean opinion score) of faithfulness.}
 \label{table:results:human:eval:faithfulness}
\end{table}

\begin{table}[t]
 \begin{center}
 \begin{tabular}{|c|c|c|}
  \hline
  \textbf{Edit} & \textbf{Model} & \textbf{Naturalness} \\
  \hline
  Refine & Auffusion & 35.14\% \\
  \hline
  Refine & MUSICGEN & \textbf{64.86\%} \\
  \hline
  Replace & Auffusion & 40.00\% \\
  \hline
  Replace & MUSICGEN & \textbf{60.00\%} \\
  \hline
  Reweight & Auffusion & 17.95\% \\
  \hline
  Reweight & MUSICGEN & \textbf{82.05\%} \\
  \hline
 \end{tabular}
\end{center}
 \caption{Comparison of audio editing capabilities of MUSICGEN an\textbf{d Auffusion based} on MOS (Mean opinion score) of naturalness.}
 \label{table:results:human:eval:naturalness}
\end{table}

\begin{table}[t]
\resizebox{\columnwidth}{!}{

 \begin{centering}
 \begin{tabular}{|c|c|c|c|}
  \hline
    \textbf{Melody} & \textbf{Dynamics} & \textbf{Tempo} & \textbf{Alignment} \\
  \hline 
  $2.66 \times 10^{-12}$ & $2.32 \times 10^{-4}$ & $2.85 \times 10^{-15}$ & $3.76 \times 10^{-06}$ \\
  \hline
 \end{tabular}
\end{centering}
}
 \caption{p-values, obtained using an unpaired t-test, indicating the statistical significance for the opinion scores obtained from the human study.}
 \label{table:results:human:eval:z_test}
\end{table}

\subsection{Human study}

\begin{figure}
 \centerline{
 \includegraphics[width=0.9\columnwidth]{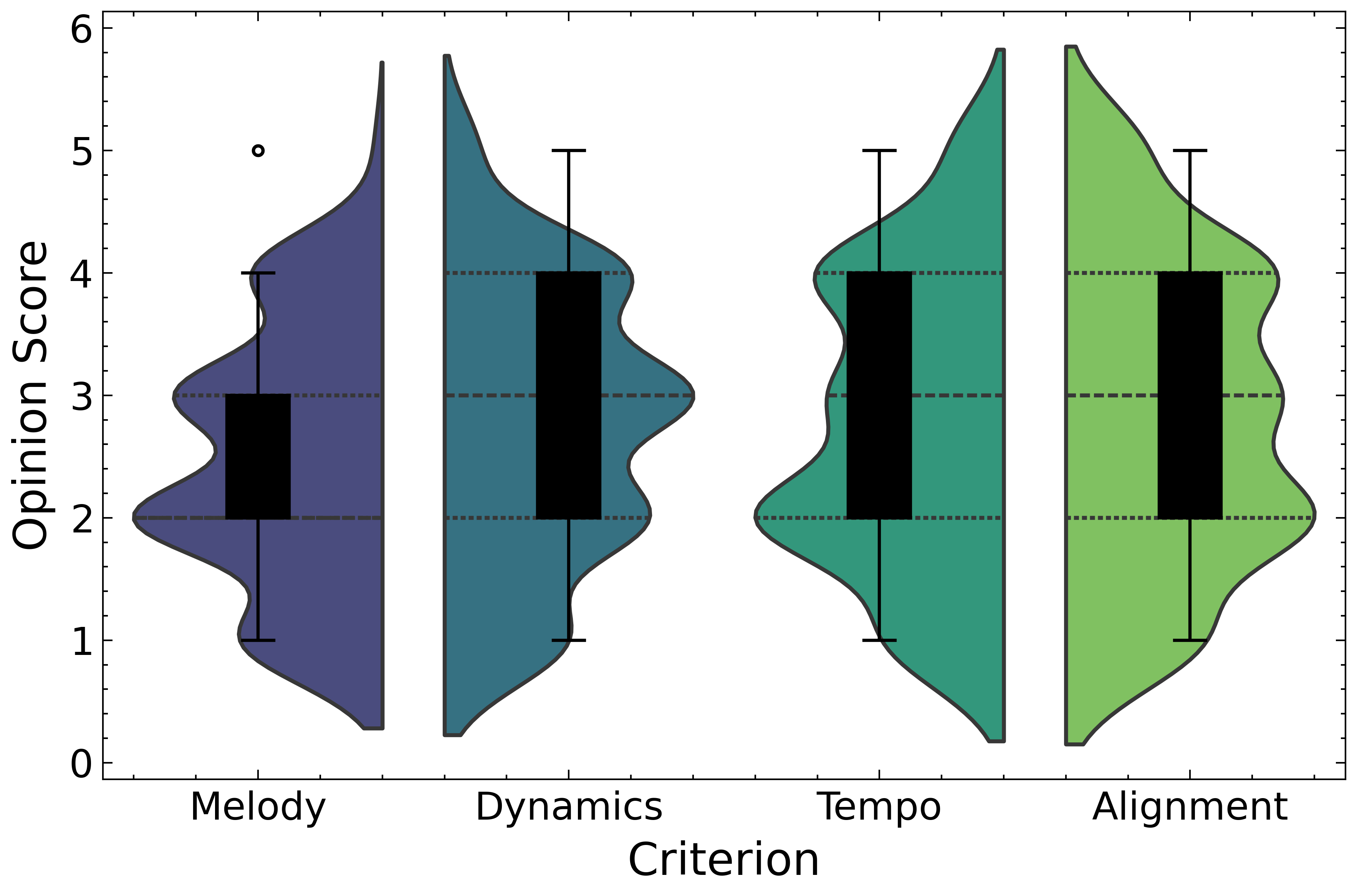}}
 \caption{Distribution of MOS (Mean opinion score) in the case of Auffusion per evaluation axis.}
 \label{fig:results:human:mos:auffusion}
\end{figure}

\begin{figure}
 \centerline{
 \includegraphics[width=0.9\columnwidth]{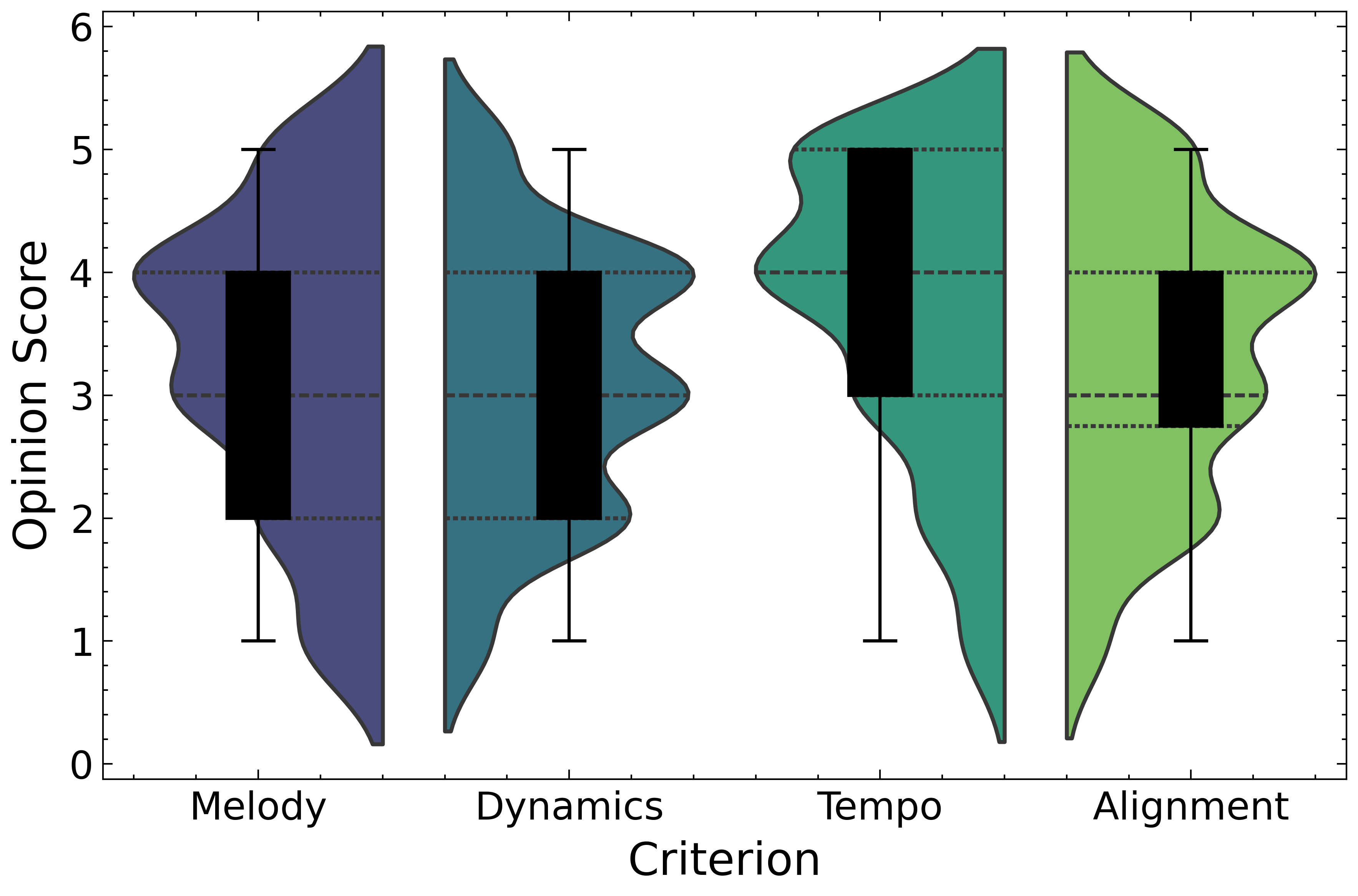}}
 \caption{Distribution of MOS (Mean opinion score) in the case of MUSICGEN per evaluation axis.}
 \label{fig:results:human:mos:musicgen}
\end{figure}

To gauge how well our method preserves naturalness and adheres to the original audio content, we invited 24 evaluators and conducted a user study. The user study began with participants evaluating pairs of 10-second audio clips. Their task was to identify the clip within each pair that exhibited the highest degree of naturalness, characterized by a resemblance to familiar musical instruments as opposed to noise or artifacts. These clips, randomly sampled from our dataset, originated from the same text prompt but were generated by distinct models – Auffusion and MUSICGEN. Next, participants assessed the fidelity of 16 randomly sampled text-audio pairings in our dataset. Each pairing included both the original text prompt and audio, alongside an edited version. Participants assessed the extent to which the edited audio remains faithful to the original version, considering key elements like melody, tempo, and dynamics, as well as its textual alignment to the edited text prompt, rating these characteristics on a Likert \cite{joshi_likert_2015} scale from 1 to 5. Based on the results of \tabref{table:results:human:eval:naturalness}, our method was found to produce audio clips perceived as more natural by participants in our study. Mean Opinion Scores (MOS) for each criterion are depicted in \figref{fig:results:human:mos:auffusion} and \figref{fig:results:human:mos:musicgen}. To assess the statistical significance of our results we employ an unpaired t-test for the opinion score distributions. \tabref{table:results:human:eval:z_test} summarizes the obtained p-values, indicating that the improvement obtained by utilizing the proposed music editing technique in conjunction with MUSICGEN significantly outperforms the baseline.


\section{Conclusion and Future works}\label{sec:summary}

In conclusion, we explored using two models for audio editing: Auffusion and MUSICGEN. We began with Auffusion, leveraging its existing capabilities for prompt-based editing. Furthermore, we introduced an alternative approach by incorporating MUSICGEN, a pre-trained auto-regressive model known for its advanced capabilities. To align Prompt-to-Prompt with MUSICGEN's auto-regressive features, we applied the attention injection procedure at all timesteps. Additionally, we introduced soft-blending, a technique that merges the feature maps generated during the forward process with the injected ones, using a weighted average for the output. This novel method aims to enhance sample quality by replicating the original diffusion-based approach of prompt-to-prompt, where edits are applied for a set number of diffusion iterations. In our evaluation, MUSICGEN outperformed Auffusion across all evaluation metrics and editing categories. It showed superior accuracy in capturing melodies, better similarity to both the original audio and target text prompt, and notably exceeded Auffusion in dynamics correlation and rhythm F1 score. This study marks the first successful application of prompt-to-prompt in the auto-regressive model audio editing context.

Looking ahead, one promising avenue is to utilize our method to construct an audio editing dataset consisting of original prompt and audio, as well as edited prompt and audio, providing valuable resources for further research and benchmarking. We also aim to conduct extensive user studies, particularly focusing on diverse ethnic music and societal contexts, which can help evaluate the effectiveness and inclusivity of the proposed audio editing techniques in various cultural settings. A limitation we faced stems from MUSICGEN's cross-attention mechanism, where all sequence information is aggregated into a single attention value. Expanding this mechanism to further investigate the interaction between text and audio tokens could produce valuable insights and yield significant improvements in audio editing quality.

\bibliography{ISMIRtemplate}

%
%
%
%
%

\end{document}